# Closed-form hard-sphere thermodynamics under nanoscale confinement: from equations of state to unmixing forces

Jose M. G. Vilar[1,2,*], and Leonor Saiz[3,†]

*[1] Biofisika Institute (CSIC, UPV/EHU), University of the Basque Country (UPV/EHU), Bilbao, Spain*

*[2] IKERBASQUE, Basque Foundation for Science, 48011 Bilbao, Spain*

*[3] Department of Biomedical Engineering, University of California, Davis, CA 95616, USA*

Equations of state accurately describe hard-sphere thermodynamics but are generally considered to fail under nanoscale confinement. We show that this failure is geometric rather than thermodynamic: a closed-form, parameter-free mapping from apparent to effective packing fraction restores the accuracy of the Carnahan–Starling equation of state and, with it, the entire analytical thermodynamic framework, including free energies, wall pressures, chemical potentials, and the large-cavity surface response. Applied to overlapping anchored hard-sphere droplets, a geometry inherent to biomolecular force generation, it yields unmixing free energies and full distance-dependent force profiles in quantitative agreement with simulations from the dilute regime to above the bulk freezing transition.

Excluded-volume interactions shape the behavior of dense fluids, colloids, and crowded biomolecular assemblies, making hard spheres a central reference model of statistical physics [1–5]. In bulk, equations of state (EOS), most notably Carnahan-Starling (CS) [6], describe hard-sphere thermodynamics with remarkable accuracy. Under nanoscale confinement, however, the space explored by particle centers deviates from the physical space occupied by particle bodies, so the apparent packing fraction ceases to be the thermodynamically relevant control parameter [7]. One is therefore generally led to abandon simple equations of state in favor of density-functional theories, simulation-based treatments, or other numerical approaches that resolve the confined system case by case [4,8–13].

Here we show, conversely, that bulk thermodynamics remains accurate once the geometric mismatch between configuration space and interaction space is treated explicitly. Separating these two spaces yields a closed-form mapping from apparent to effective packing fraction. The ideal contribution to the free energy is then governed by the center-accessible volume, while the excess contribution is evaluated at the effective packing fraction, defined as the fraction of the center-accessible volume occupied by particle bodies. This extends CS EOS to finite confined hard-sphere systems without adjustable parameters, preserving a closed-form route to confined thermodynamics.

We develop and validate this formalism in two geometries of increasing complexity. For hard spheres confined in a spherical cavity, the rescaling makes the full thermodynamic framework analytically accessible, with the resulting free-energy and wall-pressure expression closely agreeing with exact results and simulations, even at the densest packings accessible. For two overlapping anchored hard-sphere subsystems (Fig. 1), the same rescaled EOS accurately predicts unmixing free energies and the full distance-dependent force profile. This wall-free geometry is especially demanding because it simultaneously involves confinement, spatial overlap, and particle redistribution without rigid boundaries, providing a stringent test of whether the separation between configuration space and interaction space remains predictive when two confined systems interact.

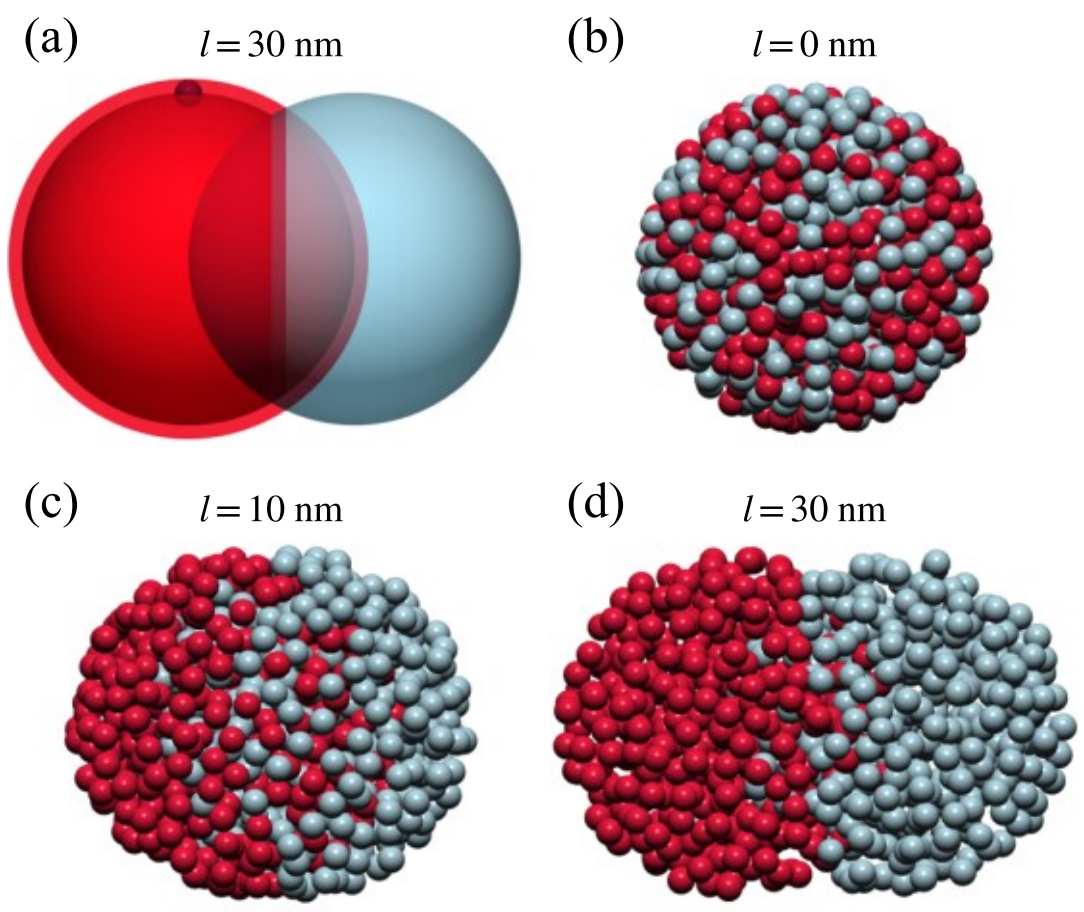

FIG. 1. Model geometry and progressive unmixing of two overlapping anchored subsystems. (a) Rendering of the model system showing two confinement volumes (translucent spheres of radius $L = 30$ nm) with their anchoring points separated by $l = L$. The faint outer shell on subsystem $a$ (red) marks the boundary layer of thickness $r = 2.5$ nm within which particle volume extends beyond the confinement region, as illustrated by a particle at the surface. (b)–(d) Renderings from Brownian-dynamics simulations of $N_s = 400$ particles per subsystem at separations $l = 0$, 10, and 30 nm, showing the progressive spatial separation of the two subsystems. Red and blue-gray spheres denote particles belonging to subsystems $a$ and $b$, respectively.

---

* Corresponding author: j.vilar@ikerbasque.org

† Corresponding author: lsaiz@ucdavis.edu

The wall-free anchored geometry is particularly relevant to biology and soft matter. Excluded-volume interactions between tethered bulky protein domains arise in multiple subcellular contexts, including membraneless condensates linked to gene regulation [14] and stress responses [15], as well as force generation between different types of macromolecular structures [16–19]. Free colloidal emulsion droplets serve as versatile platforms for confined self-assembly and entropic interactions [20–22]. In the absence of rigid walls, anchored domains can overlap, interact, and unmix through excluded-volume effects alone.

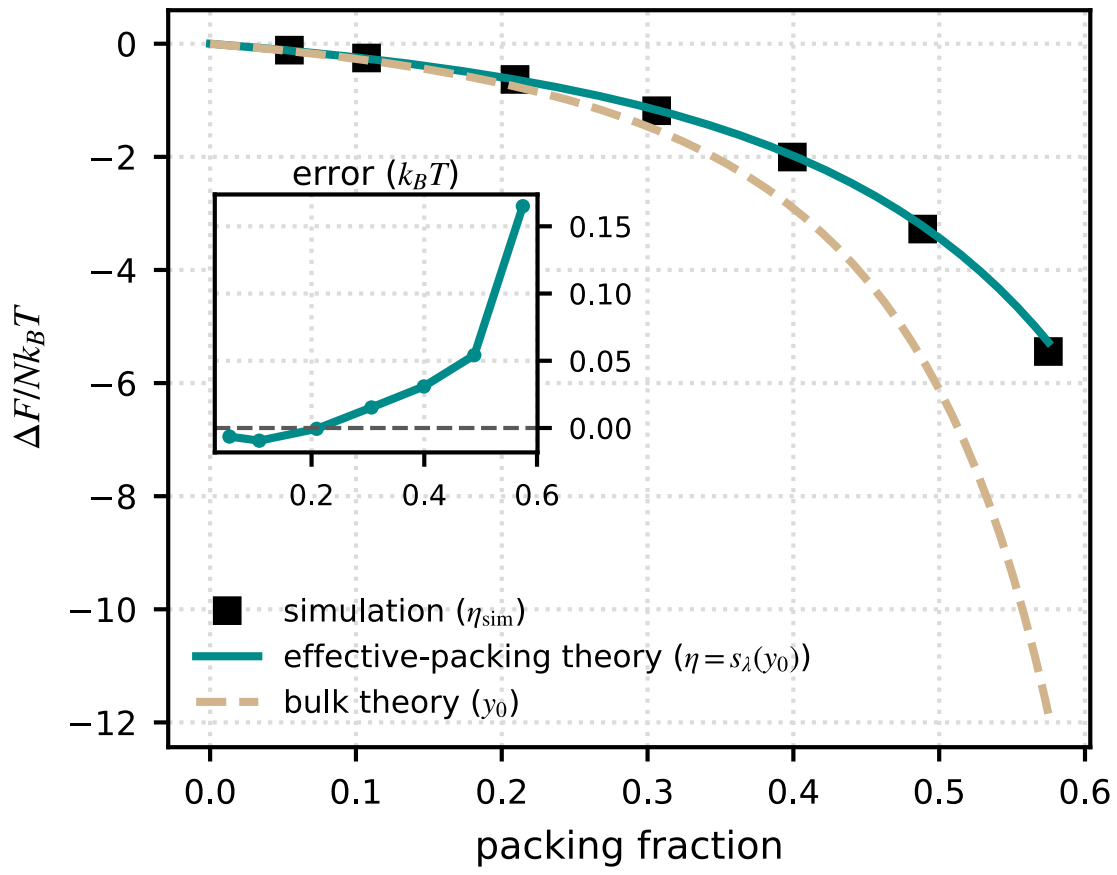

FIG. 2. Effective-packing theory recovers the free energy of unmixing. The free-energy change per particle, $\Delta F/Nk_BT$, is plotted against the packing fraction. Black squares denote simulation values plotted at the measured effective packing fraction $\eta_{\rm sim}$; the solid curve shows the effective-packing theory plotted at $\eta = s_\lambda(y_0)$; and the dashed curve shows the naive bulk theory plotted at the apparent packing fraction $y_0$. Inset: error of the effective-packing theory with respect to simulation, in units of $k_BT$ per particle.

We model this regime with two anchored hard-sphere subsystems, $a$ and $b$, each containing $N_s$ hard spheres of radius $r$. Particle centers are confined within a sphere of radius $L$ around the corresponding anchoring point, and the two anchoring points are separated by a distance $l$ (Fig. 1).

To test the theory, we performed Brownian-dynamics simulations for the anchored geometry of Fig. 1 over increasing anchoring-point separation $l$ (Appendix A). We chose $r = 2.5$ nm and $L = 30$ nm, representative of biomolecular and nanoscale systems. Starting from full overlap, $l = 0$, we increased the separation stepwise and computed the equilibrium forces $\phi_a$ and $\phi_b$ acting on the two anchoring points. For symmetric separation about the center of mass, each anchor moves by $\Delta l/2$, so the free-energy change per step is $\Delta F_l = 1/2\,[\phi_b(l) - \phi_a(l)]\,\Delta l$, and the total free-energy change is $\Delta F = \sum_{0<l<2L+2r} \Delta F_l$. The resulting $\Delta F$ is always negative and its magnitude increases sharply with density (Fig. 2), confirming that the two subsystems unmix spontaneously.

To derive the free-energy changes analytically, we take the CS EOS as the bulk hard-sphere reference, $PV/Nk_BT = (1 + \eta + \eta^2 - \eta^3)/(1-\eta)^3$, where $\eta$ is the packing fraction. In bulk, this equation of state remains highly accurate up to $\eta \simeq 0.55$. The corresponding configurational part of the Helmholtz free energy for each of the anchored subsystems is $F_s = -N_s k_B T \ln[V f_V(\eta)] + k_B T \ln N_s!$ with

$$f_V(\eta) = \exp\left[-\frac{\eta(4-3\eta)}{(1-\eta)^2}\right]. \tag{1}$$

At full overlap, both subsystems occupy the same confinement volume, and each particle effectively experiences twice the local density. Upon full separation, the density drops to the single-subsystem value. Defining the apparent packing fraction at full overlap as $y_0 = N(r/L)^3$ with $N = 2N_s$, the bulk reference is $\Delta F_{\text{naïve}} = Nk_BT\ln[f_V(y_0)/f_V(y_0/2)]$, which is always negative as unmixing is spontaneous. This gives the correct qualitative behavior but increasingly overestimates the magnitude of $\Delta F$ at higher densities (Fig. 2).

*Apparent-to-effective packing fraction mapping*— The anchoring constraint confines each particle center to a sphere of radius $L$, whereas part of the particle body can extend beyond that boundary [Fig. 1(a)]. The packing fraction based on the center-accessible volume therefore overestimates the particle volume actually inside the confinement region. We accordingly use the effective packing fraction, which considers only the particle volume contained within the center-accessible region and is the relevant variable for excluded-volume thermodynamics.

To quantify this mismatch, consider a center-accessible domain of volume $V = 4\pi L^3/3$ containing $N$ particles of volume $v_p = 4\pi r^3/3$. The apparent packing fraction is $y = v_t/V$, with $v_t = Nv_p$. Decomposing $v_t$ into contributions inside ($v_i$) and outside ($v_o$) the confinement volume, $v_t = v_i + v_o$, the effective packing fraction, $\eta = v_i/V$, can be expressed in terms of $y$ as $\eta = y/(1 + v_o/v_i)$. The geometric problem then reduces to estimating the ratio $v_o/v_i$ . We compute this ratio exactly in the dilute regime, estimate it at high density from 2D and 3D close random-packing values, and interpolate the values between these two regimes.

In the dilute regime, the protruding fraction $p_0 \equiv v_o/v_t$ equals the probability that a uniformly distributed point in the particle ball lies outside the confinement sphere. Writing this point as $\mathbf{C} + \mathbf{U}$, with $\mathbf{C}$ uniform on the center-accessible ball $B_L$ and $\boldsymbol{U}$ uniform on the particle ball $B_r$, $p_0 = \Pr(|\mathbf{C} + \mathbf{U}| > L)$. At fixed displacement $u = |\mathbf{U}|$, the fraction of centers with $\mathbf{C} + \mathbf{U}$ outside $B_L$ is $3u/4L - u^3/16L^3$, obtained from the lens volume of two equal spheres of radius $L$ separated by $u$. Averaging over $u$ with the radial density $3u^2/r^3$ of the particle ball gives $p_0(\lambda) = \frac{9}{16}\lambda - \frac{1}{32}\lambda^3$, with $\lambda = r/L$. The corresponding dilute-limit effective packing fraction is $\eta_{\rm dil} = y\,[1 - p_0(\lambda)]$. Here $p_0$ is an exact one-body occupation correction, distinct from the confined second virial coefficient, a two-body pair-exclusion quantity governed by $2r$.

At high density, we consider the boundary region as a random-packed two-dimensional surface layer with packing fraction $f_{2D}$ and the interior as a random-packed three-dimensional bulk with packing fraction $f_{3D}$. A surface layer of area $S$ contains approximately $f_{2D}S/\pi r^2$ particles and the fraction of each sphere volume lying outside is $q_{\text{out}}(\lambda) = 1/2 + 3\lambda/16$. Therefore, the protruding boundary-layer volume is $v_o \approx \frac{2}{3} S f_{2D} r\,(1 + 3\lambda/8)$, which with $v_i \approx f_{3D}V$ leads to $v_o/v_i \approx \gamma\lambda(1 + 3\lambda/8)$, where $\gamma \equiv 2f_{2D}/f_{3D}$. The analytical close random-packing values $f_{2D} = 0.886$ and $f_{3D} = 0.659$ [23] result in $\gamma = 2.69$. The corresponding dense-limit effective packing fraction is estimated as $\eta_{\text{den}} = y/[1 + \gamma\lambda(1 + 3\lambda/8)]$.

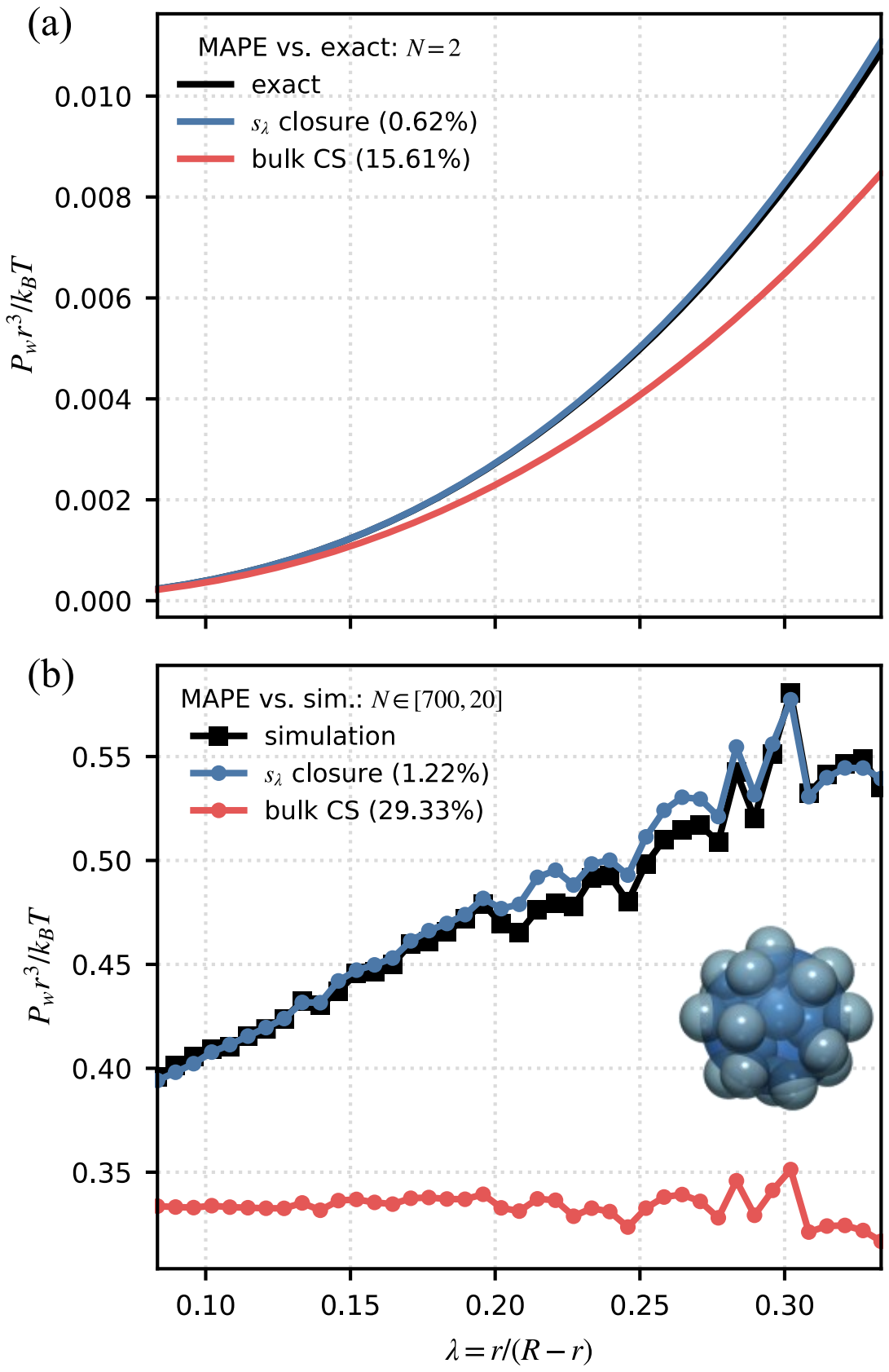

FIG. 3. Rescaled equation of state predicts wall pressure in spherical cavities. Reduced wall pressure, $P_w r^3/k_B T$, is shown as a function of the confinement ratio $\lambda$ over $\lambda \in [1/12, 1/3]$. (a) $N = 2$: exact result (black) [24] is shown alongside the $s_\lambda$ closure (blue) and the bulk CS EOS (red). (b) $N$ varies with $\lambda$ to hold the physical cavity number density $\rho = N/V_R$ approximately constant, $N(\lambda) = \text{round}[700\,V_R(\lambda)/V_R(1/12)]$, with $V_R = 4\pi R^3/3$ and $R = r(1 + \lambda^{-1})$. Simulation results (black squares) are shown alongside the $s_\lambda$ closure (blue) and the bulk CS EOS (red). Statistical errors are smaller than the symbol size. Parenthetical values in the legends give the mean absolute percentage error relative to the exact result in (a) and to simulation in (b). Bulk CS EOS evaluations use the physical-volume packing fraction $y_R = v_p\rho$. Inset: representative configuration at $N = 20$ and $\lambda = 1/3$; the translucent blue sphere marks the boundary of the center-accessible volume.

To span the full density range, we interpolate $v_o/v_i$ linearly in $\eta$ between its dilute-limit value $\alpha_0(\lambda) = p_0(\lambda)/[1 - p_0(\lambda)]$ and its high-density estimate $\alpha_*(\lambda) = \gamma\lambda(1 + 3\lambda/8)$ as $v_o/v_i = \alpha_0(\lambda) + b(\lambda)\eta$, with $b(\lambda) = [\alpha_*(\lambda) - \alpha_0(\lambda)]/f_{3D}$. This interpolation establishes a quadratic relationship $b(\lambda)\eta^2 + [1 + \alpha_0(\lambda)]\eta - y = 0$ between effective and apparent packing fractions, which leads to

$$s_\lambda(y) = \frac{\sqrt{[1 + \alpha_0(\lambda)]^2 + 4b(\lambda)y} - [1 + \alpha_0(\lambda)]}{2b(\lambda)}. \tag{2}$$

This closure recovers the exact dilute-limit relation $\eta = y\,[1 - p_0(\lambda)]$ and matches the dense boundary-layer estimate at $\eta = f_{3D}$ without adjustable parameters.

We validated $s_\lambda(y)$ against the effective packing fraction $\eta$ measured directly in simulation (Appendix A) for systems with different numbers of particles at fixed confinement radius [Fig. A1(a)] and with different confinement radii at fixed number of particles [Fig. A1(b)]. In both cases, $s_\lambda(y)$ closely follows the observed $\eta$ across the full range of packing fractions, while $y$ departs substantially from $\eta$.

Accounting for this geometric rescaling, the free-energy change becomes

$$\Delta F = N k_B \mathrm{T} \ln \frac{f_V[s_\lambda(y_0)]}{f_V[s_\lambda(y_0/2)]}. \tag{3}$$

This rescaled expression agrees closely with the Brownian-dynamics results across the full density range (Fig. 2), whereas the naive bulk theory increasingly overestimates the magnitude of unmixing at higher packing fractions.

*Spherical cavity thermodynamics*—The approach applies directly to a spherical cavity of radius $R$, a prototypical confinement geometry [25], equivalent to a single anchored system with $L = R - r$. The cavity free energy is given by $F_{\text{cav}}/k_B T = -N \ln V_{\text{acc}} - (N - 1)\ln f_V[s_\lambda(y_{\text{cav}})] + \ln N!$, where $V_{\text{acc}} = 4\pi(R - r)^3/3$ is the center-accessible volume, $y_{\text{cav}} = N v_p/V_{\text{acc}}$ is the apparent packing fraction, and $\lambda = r/(R - r)$. The $N - 1$ factor in the excess term properly accounts for the low-$N$ regime and recovers the ideal limit at $N = 1$. The wall pressure follows directly (Appendix B).

The theory agrees closely with the exact $N = 2$ result of Ref. [24] [Fig. 3(a)] and with simulations at larger $N$ across a wide range of confinement sizes [Fig. 3(b)], overcoming the marked discrepancies of bulk CS EOS. The agreement is particularly striking near λ = 1/3, where the system contains 20 particles [Fig. 3(b), inset] out of a geometric maximum of 32 [26]. Beyond λ = 1/3, finite-size packing becomes dominated by discrete structural rearrangements, and for λ > 1 the cavity accommodates only a single particle, placing λ = 1/3 as the practical limit of the testable regime.

*Large-cavity surface limit and DFT comparison*—The same finite-cavity free energy yields the leading surface response without introducing a surface coefficient as an independent input. Relative to the bulk reference at the same $N$ and $V_{\text{acc}}$, the confinement excess is $\Delta F_c = F_{\text{cav}} - F_{\text{bulk}}$, where the bulk reference is evaluated at the apparent packing fraction $y$. In the large-cavity limit, $\lambda \to 0$ at fixed $y$, the leading excess defines the surface coefficient on the center-accessible dividing surface, $\sigma_c = \lim_{\lambda\to 0} \Delta F_c/A_c$, with $A_c = 4\pi r^2\lambda^{-2}$, which leads to $r^2\sigma_c/k_BT = \lim_{\lambda\to 0}(y/4\pi\lambda)[\ln f_V(y) - \ln f_V(s_\lambda(y))]$. Thus, $\sigma_c$ is generated by the same finite-$N$ free energy that determines the cavity thermodynamics.

The theory agrees closely with the density functional theory (DFT)–based morphometric result of König, Roth, and Mecke (KRM) [27] for the physical-wall pressure at center-accessible radius $R_c = 5r$ [Fig. 4(a)]. Using chemical-potential matching to the reservoir packing fraction, the normalized $L^1$ differences are 2.22% for the EOS-matched Percus–Yevick (PY) input and 4.80% for CS. The independently derived mean-curvature coefficient, $r^2\rho_H^c = 2r^2\sigma_c/k_BT$, also agrees with the DFT result [Fig. 4(b)]. Matching and reconstruction details are given in Sec. II of the Supplemental Material.

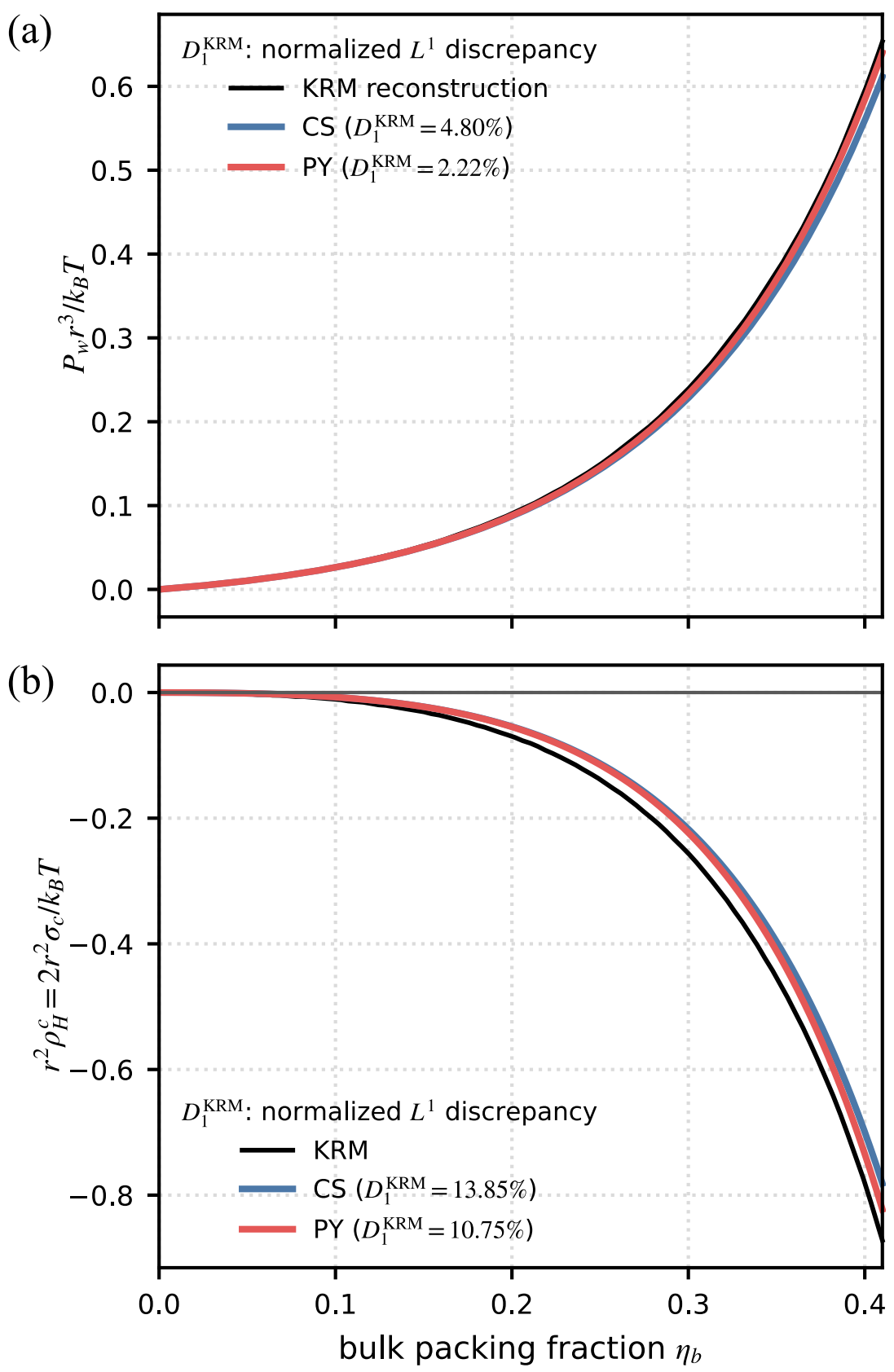

FIG. 4. Finite-curvature wall pressure and mean-curvature contact coefficient in spherical cavities. (a) Reduced physical-wall pressure $P_W r^3/k_BT$ at $R_c = 5r$ ($\lambda = 1/5$) as a function of the bulk packing fraction $\eta_b$. Black line shows the reconstruction from KRM contact-density coefficients. Blue and red lines show the theory with CS and PY EOS bulk inputs, respectively. The normalized $L^1$ differences are 4.80% for CS and 2.22% for PY. (b) Mean-curvature contact coefficient $r^2\rho_H^c = 2r^2\sigma_c/k_BT$. Black line shows the KRM result. Blue and red lines show the result from the theory with CS and PY EOS bulk inputs, respectively. Matching and reconstruction details are given in Sec. II of the Supplemental Material.

*Unmixing force profiles*—The theory extends naturally to force profiles along the separation between anchoring points $l$. At separation $l$, each confinement volume splits into an exclusive region of volume $V_e(l)$, accessible only to its own particles, and a shared region of volume $V_s(l)$, where the two confinement volumes overlap. The shared region is the lens of two spheres of radius $L$ separated by $l$, with volume $V_s(l) = \frac{\pi}{12}(2L-l)^2(4L+l)$ for $0 \le l \le 2L$ and zero otherwise. The exclusive volume is the complement, $V_e(l) = V - V_s(l)$.

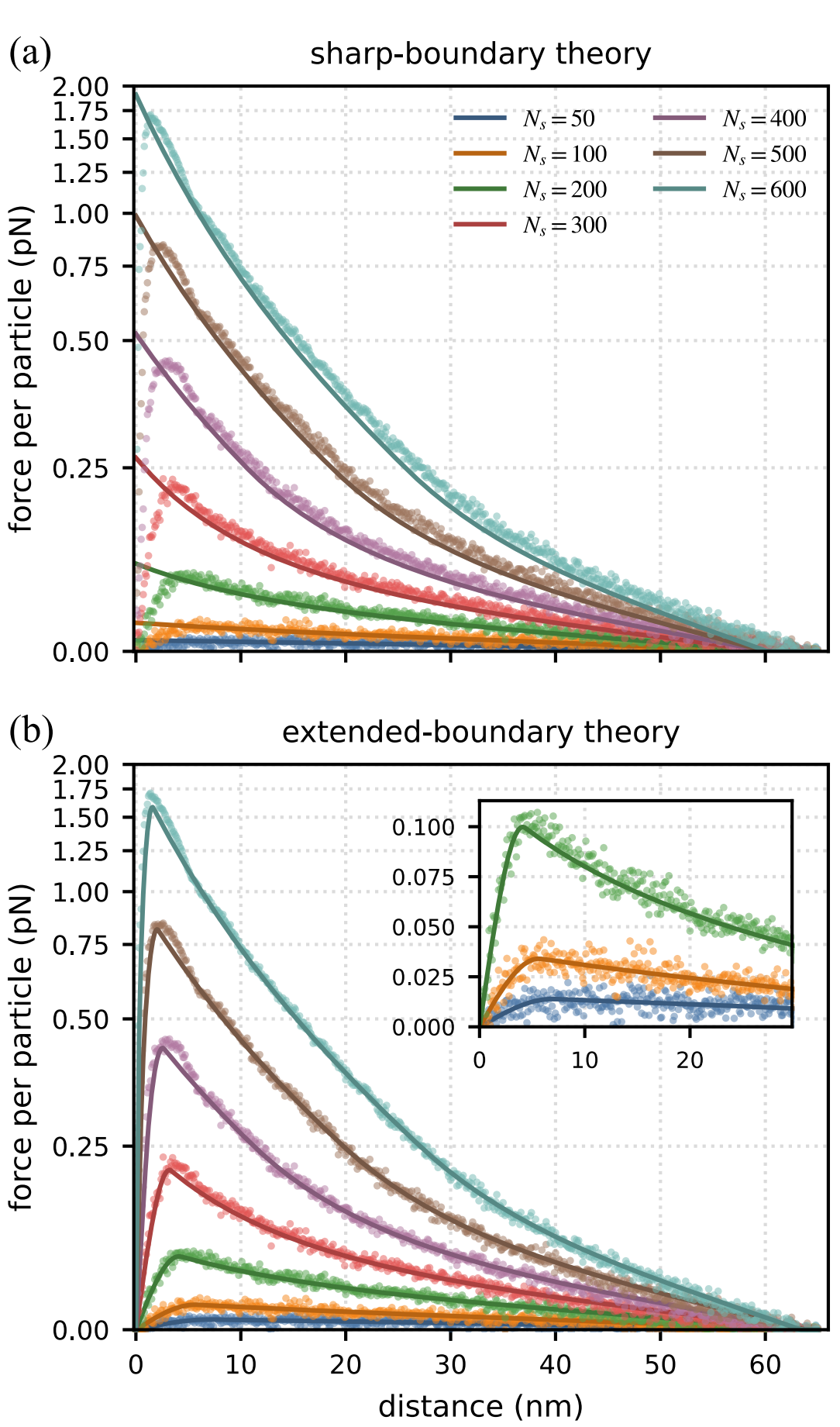

FIG. 5. Unmixing force profiles. Force per particle as a function of anchoring-point separation for systems with $N_s = 50$, 100, 200, 300, 400, 500, and 600 particles per subsystem. Dots show simulation data and solid lines show the corresponding theoretical results, with matching colors for each system size. (a) Sharp-boundary theory, with boundaries at the confinement radius $L$. (b) Extended-boundary theory, incorporating the density-dependent boundary stretching [$r_{\rm eff} = r(1 - y_0/2)$] and a matched odd cubic interpolant at short distances that preserves the total free energy change; this parameter-free construction is specified in Appendix C, Eqs. (C1)–(C4). Inset: zoom on the three lowest-density systems ($N_s = 50$, 100, 200) showing the short-distance behavior. The vertical axis uses a nonlinear scale: linear below 0.5 pN and logarithmic above.

Particles from each subsystem distribute between the exclusive and shared regions, with apparent packing fractions $y_e$ and $y_s$, respectively. The local chemical potential of a subsystem, $\mu_S = (\partial F_S/\partial N_S)_{V,T,N_{\text{ext}}}$, in the presence of particles from the other subsystem at packing fraction $y_{\text{ext}}$, is

$$\frac{\mu_S(y;y_{\text{ext}})}{k_BT} = \ln y - \frac{\partial}{\partial y}\{y \ln f_V[s_\lambda(y+y_{\text{ext}})]\}. \tag{4}$$

At each distance $l$, $y_e$ and $y_s$ are simultaneously determined by solving the equations for equal $\mu_S$ in the exclusive ($y = y_e$ and $y_{\text{ext}} = 0$) and shared ($y = y_s$ and $y_{\text{ext}} = y_s$) regions, $\mu_S(y_e;0) = \mu_S(y_s;y_s)$, and the conservation of particles, $(y_0/2)V = y_e\, V_e(l) + y_s\, V_s(l)$, in a subsystem.

Given the equilibrium partition, the free energy per particle at separation $l$ is

$$g(l) = k_BT \sum_{k\in\{e,s\}} w_k\,(l) \left\{\ln \frac{y_k}{f_V[s_\lambda(c_k y_k)]} - 1\right\}, \tag{5}$$

where $c_e = 1$ and $c_s = 2$ reflect the number of subsystems contributing to each region, and $w_k(l) = 2\,V_k(l)\,y_k/\,V y_0$ is the fraction of particles in region $k$. The excess contribution in each region is evaluated at the total local effective packing fraction $s_\lambda(c_k y_k)$.

The force per particle is $\phi(l) = -\,dg(l)/dl$, and thus combines the explicit geometric change of the overlap volume with the implicit redistribution required by chemical-potential equality.

The resulting force profiles are compared with Brownian-dynamics simulations in Fig. 5(a) for systems ranging from $N_s = 50$ to $N_s = 600$ particles per subsystem. The theory captures the overall magnitude and distance dependence of the unmixing force across all system sizes. There are two boundary effects not considered by the approach: the sharp boundary leads to a finite force at $l = 0$ but continuity and symmetry require $\phi(0) = 0$, and the force vanishes at $l = 2L$ but physical collisions persist up to $l = 2(L + r)$. Because the total free-energy change between full overlap and complete separation is fixed independently, accounting for these effects must redistribute the force profile without changing its integral.

To span the interaction range beyond the center-space boundary, we consider that particles can protrude by their full radius $r$ at low density, while crowding by the opposite subsystem reduces the typical protrusion at higher density. We therefore use the effective protrusion length $r_{\text{eff}} = r\,(1 - y_0/2)$. The force profile is then stretched to the physical range $[0, 2(L + r_{\text{eff}})]$, and its short-distance part is replaced by a matched odd cubic interpolant that enforces $\phi(0) = 0$ while preserving $\int \phi\, dl = -\,\Delta F/N$ exactly overall [Appendix C, Eqs. (C1)–(C4)]. The theory is fully determined by these conditions; no force parameter is fitted.

The extended-boundary force profiles, shown in Fig. 5(b), closely match simulation data over the full range, including the short-distance onset and resolving the two systematic deviations of the sharp-boundary profiles. Results for $L = 15$ nm ($\lambda = 1/6$), where finite-size effects are larger, show similar agreement across all system sizes studied, down to $N_s = 5$ particles per subsystem (see Supplemental Material, Figs. S1–S3).

*Discussion*—We have shown that a single geometric rescaling, independent of the underlying EOS and without adjustable parameters, extends bulk hard-sphere thermodynamics to finite confined systems. The central result is a closed-form mapping from apparent to effective packing fraction that absorbs the mismatch between center-accessible and interaction spaces. Once this mismatch is treated explicitly, the CS EOS remains quantitatively predictive under nanoscale confinement, yielding accurate parameter-free expressions for closed-form finite-$N$ canonical Helmholtz free energies, wall pressures, chemical potentials, and unmixing force profiles. For spherical cavities, the same free energy also yields the leading large-cavity surface coefficient and finite-curvature wall pressure, in agreement with independent DFT results.

The two geometries place complementary demands on the theory. The spherical cavity tests a closed-form route to thermodynamic observables across essentially the entire physically meaningful range of confinement ratios, leaving out only the magic-number regime, $\lambda \in [1/3, 1]$, where discrete cluster packing governs the physics and numerical treatment becomes unavoidable [7,10,20,25]. The overlapping anchored-droplet geometry tests a wall-free setting with inter-subsystem overlap and density-dependent particle redistribution. The same rescaled bulk EOS captures not only the total free-energy change but also the full distance-dependent force profile, particularly striking at short distances, where agreement with simulation extends well below the particle size.

Our theory provides the fixed-$N$ canonical Helmholtz free energy, $F_N = -k_\text{B}T \ln Q_N\,(V,T)$, of a closed system rather than the reservoir-state grand potential used in the DFT and morphometric descriptions considered here. The fixed-$N$ character is essential for the overlapping-domain problem, where particles redistribute between shared and exclusive regions while $N_a$ and $N_b$ remain separately conserved. The wall-free force therefore follows directly from $F(l;N_a;N_b)$ as a fixed-population interaction, not as a surface-tension or reservoir-insertion contribution, and requires neither ensemble reconstruction, constrained density-profile minimization, nor wall-based corrections. The same closed-form theory therefore describes both rigid cavities and interacting, interpenetrating domains.

The validation spans the exact $N = 2$ result, many-particle cavities with $1/12 \le \lambda \le 1/3$, an independent DFT benchmark at $\lambda = 1/5$, and overlapping populations up to $\eta_{\text{sim}} \simeq 0.575$, above bulk freezing. Across these tests, all theoretical curves in Figs. 2–5, S2, S3, and A1 are parameter-free predictions: no quantity has been fitted, and all confinement parameters are determined by geometry.

Beyond these two foundational systems, the same framework can be extended to other geometries or to bulk equations of state, including those that incorporate attractive interactions; such extensions would require separate validation of the corresponding closure. More broadly, the results identify a compact analytical route to confinement-induced free energies, pressures, and forces in crowded nanoscale systems, quantities that have until now generally required either numerical treatment or independently specified surface and curvature contributions [5].

*Acknowledgments*—J.M.G.V. acknowledges support from Ministerio de Ciencia, Innovación y Universidades (Grant PID2024-160016NB-I00 funded by MICIU/AEI/10.13039/501100011033 and by ERDF/EU).

## Appendices

*Appendix A: Computational methods*—We performed Brownian-dynamics simulations as described in Refs. [18,19]. Explicitly, the displacement $\Delta\mathbf{r}_i$ of a particle $i$ in a time interval $\Delta t$ is given by

$$\Delta\mathbf{r}_i = \frac{D}{k_B T}\mathbf{F}_i\Delta t + \sqrt{2D\Delta t}\,\mathbf{G}_i, \tag{A1}$$

where the subscript $i$ refers to the particle index, $D$ is the translational diffusion coefficient of the particle, $\mathbf{F}_i$ is the force acting on the particle, $k_B$ is the Boltzmann constant, $T$ is the absolute temperature, and $\mathbf{G}_i$ is a standard Gaussian random vector. The Stokes-Einstein relation, $D = k_B T/6\pi\mu r$, where $\mu$ is the solvent viscosity, leads to values of $D = 0.098$ nm$^2$ ns$^{-1}$ for $r = 2.5$ nm spheres in water at 25°C.

The force acting on particle $i$ comprises hard-core and anchoring interactions. Hard-core interactions between particles $i$ and $j$ have radial symmetry and are described mathematically through $\mathbf{f}_{ij} = H\left(2r - \| \mathbf{r}_i - \mathbf{r}_j \|\right)\frac{\mathbf{r}_i - \mathbf{r}_j}{\|\mathbf{r}_i - \mathbf{r}_j\|}$, where $H(x)$ is zero for $x < 0$ and infinity otherwise. The anchoring force restricts how far particles can be from their anchoring points and is described through $\mathbf{m}_i = -H(\| \mathbf{r}_i - \mathbf{a}_i \| - L)\frac{\mathbf{r}_i - \mathbf{a}_i}{\|\mathbf{r}_i - \mathbf{a}_i\|}$, where $L$ represents the confinement radius linking particle $i$ to anchoring point at position $\mathbf{a}_i \in \{\mathbf{a}_a, \mathbf{a}_b\}$. Here, $\mathbf{a}_a$ and $\mathbf{a}_b$ are the anchoring positions of subsystem $a$ and $b$, respectively.

To implement the simulations, the function $H(r)$ is approximated as $H(r) = F_W\Theta(r)$, where $\Theta$ is the Heaviside unit step function and $F_W$ is the force intensity.

The initial randomly oriented distribution of particles attached to each anchoring point was equilibrated over approximately 13 µs. To facilitate equilibration, the force intensities were increased quadratically in time during this interval from $F_W \simeq 0.01$ pN to $F_W = 100$ pN. Concomitantly, the time step was decreased quadratically from $\Delta t = 1.0$ ns to $\Delta t = 0.01$ ns. The production values were $F_W = 100$ pN and $\Delta t = 0.01$ ns.

The use of a large but finite force effectively increases the confinement radius by $\Delta L = k_B T/F_W$ and decreases the particle radius by $\Delta r = k_B T/2F_W$, which we compensated for with $r_{\mathrm{sim}} = r + \Delta r$ and $L_{\mathrm{sim}} = L - \Delta L$ in the simulations.

The force acting on the anchoring points is computed as the time average of $-\sum_i \mathbf{m}_i$. Starting at zero separation, we increased the distance by 0.1 nm, re-equilibrated the system using the same quadratic force and time-step ramp over approximately 2 µs, and then averaged the forces $\phi_a$ and $\phi_b$ at the two anchoring points over 100 ns at production conditions. This process was iterated to reach the maximum separation of $2L + 2r$. To reduce the statistical fluctuations in the computed force profile to the same level in all the cases, we averaged over replicas of the system to reach a constant total of 4800 particles, or just below, across all replicas, ranging from 3 replicas for the largest systems to 48 replicas for the smallest systems.

(a)

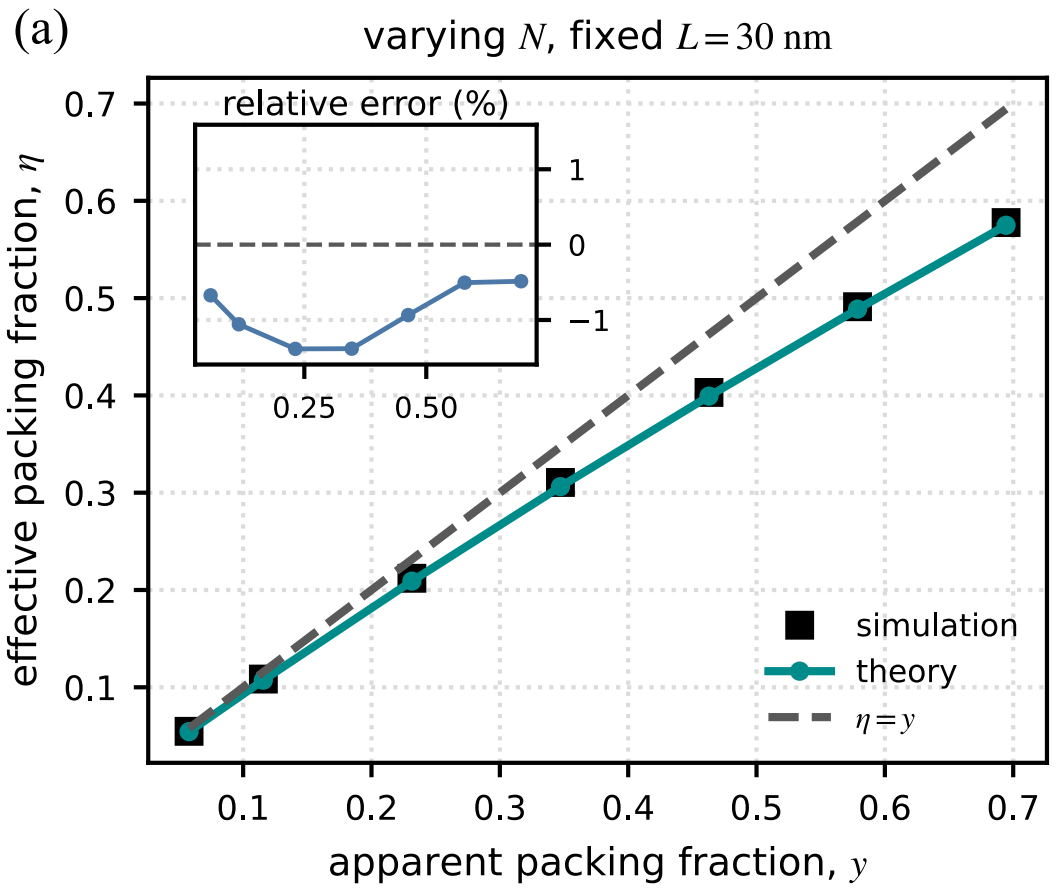

(b)

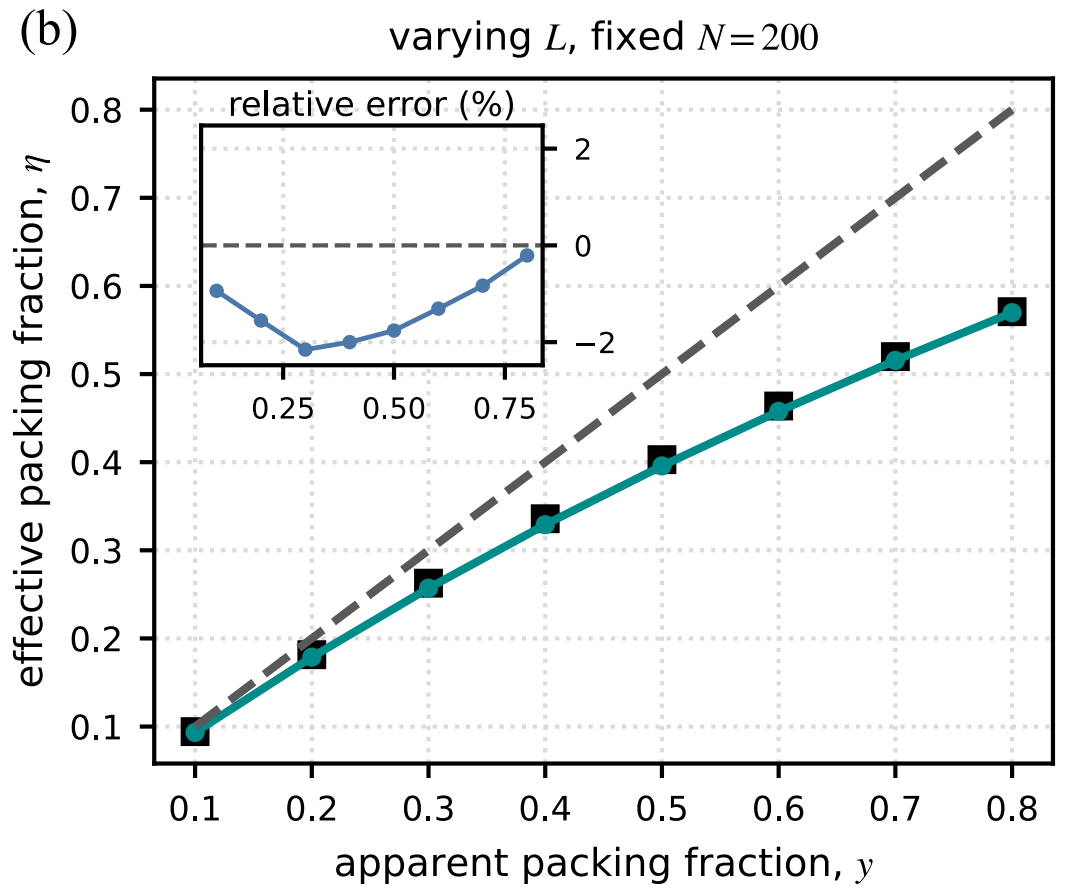

FIG. A1. Validation of the apparent-to-effective packing fraction mapping. (a) Effective packing fraction, $\eta$, versus apparent packing fraction, $y$, for systems with varying number of particles $N$ at fixed confinement radius $L = 30$ nm. (b) Same comparison for systems with varying confinement radius $L$ at fixed $N = 200$ particles per subsystem. Black squares show values extracted from simulation, the continuous cyan lines show the theoretical result $s_\lambda(y)$, and the dashed gray lines mark the naive reference $\eta = y$. Insets report the relative error of the theoretical result with respect to simulation.

The effective packing fraction $\eta$ is computed by Monte Carlo integration. Random points are sampled uniformly inside each particle's body and the fraction lying within the center-accessible region is counted. Averaging over time samples and replicas gives the simulation values plotted in Fig. A1.

The simulations for the cavity were performed using the same setup but with anchor points far apart beyond $2(L + r)$

so that subsystems do not interact. The wall pressure was computed from the radial projection of the anchoring forces via the virial relation for a spherical container:

$$P_w = -\frac{R}{3V_R}\left\langle \sum_i \mathbf{m}_i \cdot \hat{\mathbf{n}}_i \right\rangle,$$

(A2)

where $V_R = \frac{4}{3}\pi R^3$ is the physical cavity volume, $\hat{\mathbf{n}}_i$ is the radial unit vector from the anchor to particle $i$, and $\mathbf{m}_i$ is the anchoring force.

*Appendix B: Equation of state for a spherical cavity*—The confinement ratio is $\lambda = r/(R-r)$, and the apparent packing fraction based on the center-accessible volume is $y_{\text{cav}} = Nv_p/V_{\text{acc}}$, with $V_{\text{acc}} = 4\pi(R-r)^3/3$. The effective packing fraction entering the Carnahan-Starling free energy is $\eta = s_\lambda(y_{\text{cav}})$, using the same scaling function $s_\lambda$ derived in the main text.

The pressure exerted on the cavity wall, obtained from the free energy as $P_w = -(1/4\pi R^2)\,\partial F_{\text{cav}}/\partial R$, is

$$P_w = \frac{k_B T N}{4\pi R^2 L}\left[3 + \left(1 - \frac{1}{N}\right)\frac{4-2\eta}{(1-\eta)^3}\,\Xi(\lambda, y_{\text{cav}})\right],$$

$$\Xi(\lambda, y) \equiv -L\frac{\partial \eta}{\partial R} = \frac{3y - \lambda[\alpha_0'(\lambda)\eta + b'(\lambda)\eta^2]}{1 + \alpha_0(\lambda) + 2b(\lambda)\eta},$$

(B1)

where $\Xi$ collects the chain-rule contributions from the dependence of $s_\lambda$ on both $y$ and $\lambda$. This expression reduces to CS EOS only for both $\lambda \to 0$ and large $N$.

*Appendix C: Boundary refinement of the force profile*—The sharp-boundary force profile $\phi_0(l)$, derived from the center-space theory, exhibits two minor systematic deviations: it predicts a finite force at $l = 0$ (where odd symmetry requires $\phi = 0$) and vanishes at $l = 2L$ (rather than at $l = 2(L + r)$, where physical inter-subsystem collisions cease). The method below stretches the profile to the physical range and restores the symmetry constraint, while preserving the total free-energy change $\int \phi\, dl = -\Delta F/N$ exactly and introducing no adjustable parameters.

*Stretching.* The separation coordinate is rescaled so that the force profile extends to $[0, 2(L + r_{\text{eff}})]$:

$$\phi_{\text{str}}(l) = \phi_0\left(\frac{L}{L + r_{\text{eff}}}\, l\right),$$

(C1)

where $r_{\text{eff}} = r\,(1 - y_0/2)$ as defined in the main text. This maps $l = 2(L + r_{\text{eff}})$ to the argument $2L$, where $\phi_0 = 0$, so the stretched force vanishes at the correct range. The stretching increases the integral of the force by a factor $(L + r_{\text{eff}})/L$, producing an excess

$$I_{\text{excess}} = \frac{r_{\text{eff}}}{L}\, I_{\text{orig}},$$

(C2)

where $I_{\text{orig}} = \int_0^{2L} \phi_0\,(l)\, dl$.

*Short-distance matching.* Since the free energy $g(l) = g(-l)$ is even in the separation, the force $\phi = -dg/dl$ is odd: $\phi(-l) = -\phi(l)$, which requires $\phi(0) = 0$. To enforce this while absorbing the excess integral, we replace $\phi_{\text{str}}$ on the interval $[0, l^*]$ with an odd cubic interpolant:

$$\phi_{\text{ext}}(l) = \begin{cases} a_1 l + a_3 l^3 & 0 \le l \le l^*, \\ \phi_{\text{str}}(l) & l > l^*. \end{cases}$$

(C3)

The absence of even powers ensures $\phi_{\text{ext}}(0) = 0$ and preserves the odd symmetry. Requiring continuity and differentiability at $l^*$ fixes the two coefficients at $a_1 = [3\,\phi_{\text{str}}(l^*) - l^*\,\phi'_{\text{str}}(l^*)]/2\,l^*$ and $a_3 = [l^*\,\phi'_{\text{str}}(l^*) - \phi_{\text{str}}(l^*)]/2\,l^{*3}$. The remaining parameter $l^*$ is fixed by the condition that the area removed equals the excess:

$$\int_0^{l^*} [\phi_{\text{str}}(l) - \phi_{\text{ext}}(l)]\, dl = I_{\text{excess}}.$$

(C4)

The resulting profile $\phi_{\text{ext}}(l)$ is continuous and differentiable at $l^*$, vanishes at both $l = 0$ and $l = 2(L + r_{\text{eff}})$, and preserves $\int \phi\, dl = -\Delta F/N$ exactly. All quantities — $r_{\text{eff}}$, $a_1$, $a_3$, and $l^*$ — are determined by $s_\lambda$ and the matching conditions, with no adjustable parameters.